\documentclass[12pt,authoryear]{elsarticle}

\usepackage{amssymb}
\usepackage{amsmath}
\DeclareMathOperator{\D}{\mathrm{d\!}}

\usepackage{color}

\begin{document}
\begin{frontmatter}
	
\title{On phenomenology of physical effects in axons}
\author[1,2]{Jüri Engelbrecht}
\ead{je@ioc.ee}
\author[1]{Kert Tamm}
\ead{kert.tamm@taltech.ee}
\author[1]{Tanel Peets}
\ead{tanel.peets@taltech.ee}

\affiliation[1]{organization={Department of Cybernetics, Tallinn University of Technology},
            addressline={Ehitajate Rd. 5}, 
            city={Tallinn},
            postcode={19086}, 
            country={Estonia}}
\affiliation[2]{organization={Estonian Academy of Sciences},
            addressline={Kohtu 6}, 
            city={Tallinn},
            postcode={10130}, 
            country={Estonia}}
            
\begin{abstract}
This paper deals with the mathematical modelling of signal propagation in nerve fibres. Due to the complexity of the processes where electrical, mechanical, and thermal effects are coupled, a phenomenological approach helps to build mathematical models. The ideas of phenomenology are briefly presented, and their application is described. These applications cover the modelling of ion currents (the Hodgkin–Huxley model), temperature effects, and inductance. This means that the ion currents through the biomembrane, the influence of endo- and exothermic reactions on temperature, and the influence of energy in a non-electrical form are taken into account using phenomenological variables, i.e., observables. Such an approach brings the mathematical models closer to reality. Using the concept of phenomenological inductance helps us better understand the propagation of an action potential in myelinated axons. In principle, contemporary mathematical models describing the process in axons are hybrid in nature, combining physical laws with phenomenology, i.e., with observables.
            	
\end{abstract}

\begin{keyword}
phenomenology \sep axons\sep ion currents \sep temperature \sep inductance	
\end{keyword}
            
\end{frontmatter}

\section{Introduction}

Contemporary studies of biological processes are characterised by ``integrative biological modelling'' \citep{McCulloch2002} over several time and space scales.  Noble \citeyearpar{Noble2002} has stressed the need to distinguish descriptive, integrative and explanatory (predictive) levels of modelling phenomena, which, without any doubt, are extremely complex. It means that for understanding the biological processes, an interdisciplinary approach is needed, taking into account not only biological phenomena but also physics and chemistry, using mathematical descriptions. The structural elements and processes are coupled, and special attention must be paid to understand the possible interactions between them. Although it is said that physics rules biology \citep{Pennycuick1992}, due to the myriad of structural details and possible interactions of phenomena, not all the processes are presently understood in the needed detail. Nevertheless, the joint efforts of theoreticians and experimentalists have step by step deepened the understanding of how biological processes work \citep{Bialek2018,Schneider2021}.

In what follows, the attention is focused on the mathematical modelling of signal propagation in nerve fibres where electrical signals are accompanied by mechanical and thermal phenomena. The aim is to understand the interaction effects between the various phenomena and demonstrate the importance of using observables in the modelling of signals in nerves, which can be interpreted as an ensemble of waves \citep{Engelbrecht2021a}. 

\section{Background in brief}
\subsection{The structure of axon}

A simple axon, like in classical experiments of Hodgkin and Huxley \citeyearpar{Hodgkin1952}, can be modelled as a tube (in terms of mechanics) embedded in extracellular fluid. Inside the tube is the axoplasmic fluid (intracellular fluid or axoplasm). The wall of the tube forms a barrier between the extracellular and intracellular fluids. This barrier is a lipid biomembrane with proteins \citep{Mueller2014}, but often it is modelled as a lipid bilayer only \citep{Heimburg2005,Engelbrecht2021a}. The main signal propagating in an axon is the action potential (AP). The signal formation and propagation are governed by the changes in the concentration of ions in the axoplasm. The ion currents which cause these changes occur through ion channels in the barrier. It means that in structural terms, several types of ions (mostly Na$^+$ and K$^+$) can pass through ion channels from the axoplasm into the extracellular fluid and vice versa. 
In case of myelinated axons, the structure is more complicated. Namely, the biomembrane (a lipid bilayer) is surrounded by a myelin sheath composed of multiple layers of lipid bilayers \cite{Poitelon2020}. Under the myelin sheath, the ion currents are suppressed. However, an axon is not fully covered by the myelin sheath; it is interrupted by nodes of Ranvier, where there is a high concentration of sodium channels that can pass the ions. Potassium channels are concentrated in the regions adjacent to the nodes of Ranvier. Such a structure certainly influences the properties (especially the velocity) of an AP. The complexity of the signal propagation in both cases (unmyelinated and myelinated axons) is characterised by accompanying mechanical and thermal effects. 
 
This brief description is elaborated in much more detail in many studies (see, for example,  Clay \citeyearpar{Clay2005}; Debanne et al. \citeyearpar{Debanne2011}).

\subsection{Physical processes}

The understanding of physical processes in nerves is based on numerous experiments. There are several components of the wave ensemble that should be measured:\\
(i) the action potential AP;\\
(ii) the longitudinal wave LW in the biomembrane;\\
(iii) the transverse wave TW in the biomembrane;\\
(iv) the pressure wave PW in the axoplasm;\\
(v) the temperature $\Theta$. \\
Besides measuring the AP  (see Hodgkin and Huxley, \citeyear{Hodgkin1945}), the other effects (mechanical and thermal) are measured in many experiments \citep[among others]{Abbott1958,Howarth1968,Iwasa1980a,Terakawa1985,Tasaki1989,Yang2018}. Up to now, it seems that although LW and TW are physically coupled, there are no direct measurements of LW. 

The processes to be understood are the following:\\
(i) the formation and generation of an AP and its dependence on ion currents;\\
(ii) electric-biomembrane interaction resulting in mechanical waves LW and TW in the biomembrane;\\
(iii) electric-fluid (axoplasm) interaction resulting in a mechanical wave PW in the biomembrane;\\
(iv) electric-fluid (axoplasm) interaction resulting in the thermal response in the axon.

If possible, then also possible feedback should be analysed. Leaving aside the details, a possible general model for an unmyelinated axon based on accounting of processes listed above is proposed by Engelbrecht et al. \citeyearpar{Engelbrecht2021a} and compared with other mathematical models by Peets et al. \citeyearpar{Peets2023}. The coupling between the components of the wave ensemble is described by interaction  (coupling) forces \citep{Engelbrecht2021a} based on the analysis of physical mechanisms of interaction. The main hypothesis is \citep{EngelbrechtMEDHYP}:
 all the mechanical effects in the axoplasm and the surrounding biomembrane, together with the heat production, are generated due to changes in electrical signals (AP or ion currents) that dictate the functional shape of coupling forces.

\section{Modelling and phenomenology}

Mathematical modelling is a widely used tool for describing physical phenomena. It is said sometimes that modelling means understanding. The backbone of modelling is based on using the laws of physics. However, it means that all the physical mechanisms are described in detail, and in the case of many coupled processes, all mechanisms of coupling are understood and also described with sufficient accuracy. In the case of such complex biological processes as the propagation of signals in nerves, the modelling is a real challenge. Andersen et al. \citeyearpar{Andersen2009} stressed that there is a need ``... to frame a theory that incorporates all observed phenomena in one coherent and predictive theory of nerve signal propagation.'' Later, this idea is elaborated in many studies (see, for example, an overview of models by Peets et al., \citeyear{Peets2023}). Nevertheless, as mentioned by Drukarch et al. \citeyear{Drukarch2018}, the studies for ``... the formulation of a more extensive and complete conception of the nerve impulse'' should continue. 

The reason why one should continue research towards ``a more extensive and complete conception'' is explained by Engelbrecht et al. \citeyear{Engelbrecht2024}:                
``It is not always possible to understand the essence of physical mechanisms, especially when several of them are coupled. Nevertheless, observations (experiments) help to understand many aspects of processes under investigation, which permit to describe empirical relationships between the phenomena. In this case, a model is based on phenomenology rather than on physical theory. Portides \citeyearpar{Portides2011} explains that such a model compensates for the lack of knowledge `of how exactly and to what extent each part of a system contributes to the latter’s investigated behaviour'. In other words, most of the system is too complex for a straightforward application of fundamental laws, which are typically composed for an idealised model system, and phenomenological models are used to describe the relationships between the variables of the model within the measured values.''

The ideas of using phenomenology in the philosophy of science are based on the studies of Edmund Husserl (1859-1938). His ideas to use phenomena (observations) rather than explanations have turned very productive \citep{Heelan1991}. A brilliant example of using phenomenology was proposed by Heisenberg \citeyearpar{Heisenberg1925}. Instead of looking for the description of the motion of electrons from one orbit to another, he proposed to use observables and constructed the matrices for stating only the initial and final positions of electrons. This idea - instead of using variables to use tables with positions - was crucial in developing the quantum theory \cite[see][]{Rovelli2022}.

In this way, in addition to science-driven models based on using the physical laws for describing the processes, the usage of phenomenological models based on observables might give more information about the phenomenon under investigation.  Morrison \citeyearpar{Morrison1999} states that many models ``have a rather hybrid nature'' - neither purely science-driven nor phenomenological. In what follows, we focus our attention on the modelling of signal propagation in nerves and describe the hybrid models, starting with the classical Hodgkin-Huxley (HH) model up to modelling of heat production in nerves and the role of inductance, which is a hot topic in neuroscience \citep{Wang2021}. Finally, we discuss how phenomenology may help to model the processes in myelinated nerves better.

\section{Ion currents}

The celebrated Hodgkin-Huxley (HH) model is a cornerstone of neuroscience \citep{Hodgkin1952,Hodgkin1964}. Based on their experiments on the unmyelinated axon of a giant squid, they proposed a hybrid model for the AP involving potential difference $V$ across the biomembrane and the membrane current I as a driving component. The membrane current (per unit area) is calculated as
\begin{equation}
	I=C_M\frac{\D V}{\D t}+I_{\mathrm{Na}}+I_{\mathrm{K}}+I_{\mathrm{L}},
\end{equation}
where $C_M$ is the membrane capacitance and $I_\mathrm{Na}$, $i_\mathrm{K}$ and $I_\mathrm{L}$ are sodium, potassium and leakage currents respectively. 

The ingenious proposal of Hodgkin and Huxley \citeyearpar{Hodgkin1952} was to introduce phenomenological variables $n$, $m$ and $h$ related to $I_\mathrm{K}$ and $I_\mathrm{Na}$. Why and how they came to such an idea is commented on by Raman and Ferster \citeyearpar{Raman2021}. For the potassium conductance $g_\mathrm{K}$, they proposed:
\begin{align}
	g_\mathrm{K}&=n^4\bar{g}_\mathrm{K},\\
	\frac{\D n}{\D t}&=\alpha_n(1-n)-\beta_n n,
\end{align}
where $\alpha_n$ and $\beta_n$ are rate voltage-dependent rate coefficients characterising opening and closing of the gate respectively; $\hat{g}_\mathrm{K}$ is the maximum potassium conductance. They explained \citep{Hodgkin1952}: ``These equations may be given a physical basis if we assume that potassium ions can only cross the membrane when four similar particles occupy a certain region of the membrane. $n$ represents the proportion of the particles in a certain position (for example, at the inside of the membrane) and $1-n$ represents the proportion that are somewhere else (for example, at the outside of the membrane). $\alpha_n$  determines the rate of transfer from outside to inside, while $\beta_n$ determines the transfer in the opposite direction.''

The description of sodium conductance needs activation (internal variable $m$) and inactivation (internal variable $h$) to be taken into account. It was proposed:
\begin{align}
	\label{HHnmhEvol1}
	g_\mathrm{Na}&=m^3h \bar{g}_\mathrm{Na},\\
	\label{HHnmhEvol2}
	\frac{\D m}{\D t}&=\alpha_m(1-m)-\beta_m m,\\
	\label{HHnmhEvol3}
	\frac{\D h}{\D t}&=\alpha_h(1-h)-\beta_h h. 
\end{align}
Here also the rate coefficients $\alpha$ and $\beta$ characterise opening and closing of the ion channels and $\hat{g}_\mathrm{Na}$ is the maximum sodium conductance. The explanation is the following \cite{Hodgkin1952}: ``These equations may be given a physical basis if sodium conductance is assumed proportional to the number of sites on the inside of the membrane which are occupied simultaneously by three activating molecules but are not blocked by an inactivating molecule. $m$ then represents the proportion of activating molecules on the inside and $1-m$ the proportion on the outside; $h$ is the proportion of inactivating molecules on the outside and $1-h$ the proportion on the inside.''

The membrane current (per unit area) is then calculated as:
\begin{equation}
\label{MembCurrent}
	I=C_M\frac{\D V}{\D t}=g_\mathrm{K}n^4(V-V_\mathrm{K})+g_\mathrm{Na}m^3h(V-V_\mathrm{Na})+g_\mathrm{L}(V-V_\mathrm{L}),
\end{equation}
where $g_\mathrm{L}$ is leakage conductance and $V_\mathrm{K}$, $V_\mathrm{Na}$ and $V_\mathrm{L}$ are respective equilibrium (Nernst) voltages. 

Note that the phenomenological variables in Eq.~\eqref{MembCurrent} are in a combination ($n$ and $mh$) which is related to physical phenomena according to the quotations above. The rate constants were determined by the careful fitting of theoretical calculations to the experiments. 

Without any doubt, the HH model is an excellent example of how physical and phenomenological variables together permit to describe the phenomenon.

\section{Temperature}

Experimental studies have demonstrated that the propagation of an AP is accompanied by the generation and absorption of heat and, consequently, by temperature changes \citep{Abbott1958,Ritchie1985,Tasaki1988}.  The possible mechanisms of heat production can be explained by Joule heating (energy transfer from the electrical current to thermal effects) and also by energy transfer from mechanical waves in the biomembrane and axoplasm into the temperature increase. In addition, Abbott et al. \citeyearpar{Abbott1958} proposed that ``... the positive heat is due to exothermic chemical reactions ... and the negative heat to endothermic reactions''. This proposal can be treated as a possible explanation or as a hypothesis. Tamm et al. \citeyearpar{Tamm2019} have proposed to use phenomenological approach to describe this phenomenon. In this case, the idea stems from continuum mechanics where the notion of `internal variables' is preferred to `phenomenological variables' (see Engelbrecht et al., \citeyear{Engelbrecht2021a}).

In principle, the temperature change in an axon, following the general thermodynamics, is governed by the classical heat equation with a driving force, which describes the interaction with other fields:
\begin{equation}
	\Theta_T =\alpha \Theta_{XX}+F(V,J,P,U),
\end{equation}
where $\Theta$ is temperature, $\alpha$ is thermal diffusivity, $F(V,J,P,U)$ is coupling force due to AP ($V$), generalised ion current ($J$), pressure wave in axoplasm ($P$), longitudinal wave in biomembrane ($U$)  and subscripts $T$ and $X$ denote partial derivatives wth respect to time and space respectively. 

Tamm et al. \citeyearpar{Tamm2019} have proposed that the driving force $F_\mathrm{chem}$ that is related to the chemical process accompanying the propagation of an AP, has a simple form:
\begin{equation}
	\label{F_chem}
	F_\mathrm{chem}=-\tau \Omega,
\end{equation}
where $\tau$ is a physical coefficient and $\Omega$ is an internal variable. This variable is governed by
\begin{equation}
	\label{OmegaEvol}
	\Omega_T+\varepsilon\Omega=\xi J,
\end{equation}
where $\varepsilon$, $\xi$ are coefficients and $J$ is ionic current from the model governing the AP. Note that Eq.~\eqref{OmegaEvol} is similar to Eqs.~\eqref{HHnmhEvol1}-\eqref{HHnmhEvol3} used by Hodgkin and Huxley  \citeyearpar{Hodgkin1952} for describing the ion currents. Fitting the coefficients of Eqs.~\eqref{F_chem},\eqref{OmegaEvol} against experimental results \citep{Abbott1958}, the numerically obtained results matched the pattern of temperature changes rather well \citep{Peets2021}.

\section{Inductance}

Although the starting point of Hodgkin and Huxley to derive their celebrated model with ion currents was the Maxwell equations, the inductance was neglected. Indeed, later Scott \citeyearpar{Scott1975} {\color{red}(1971)} concluded that the possible inductance in the axoplasm is too small to influence the conduction velocity. 

However, the question about the influence has been raised in many studies even before the HH model was derived. This was a hard question. Cole and Baker \citeyearpar{ColeBaker1941}: ``The concept of an inductance in a cell membrane is so foreign to our past experience and so difficult to grasp that we must inquire closely into each of the steps which have led to it before we can resign ourselves to the necessity of accepting and using it.''  For example, Cole \citeyearpar{Cole1941} noted that the inductive behaviour is not necessarily nor uniquely explained by the ability of a system to store magnetic energy, but rather by storing energy in a non-electrical form. Certainly, Hodgkin and Huxley knew about Cole's studies and his ideas about the role of inductance. However, they were sceptical about the influence of inductance \citep{Hodgkin1945}: ``We are reluctant to accept the idea of a genuine inductance in the membrane, since it is difficult to attach any physical significance to this concept.'' Nevertheless, later they mentioned  \citep{Hodgkin1952} that ``in our theoretical model the inductance is due partly to the inactivation process and partly to the change of potassium conductance, the latter being somewhat more important''. They also estimated the value of inductance as 0.39~H at 25~°C, which is of the same order as 0.2~H found by Cole and Baker \citeyearpar{ColeBaker1941}. It seems, however, that the scientific community compared nerve fibres with wires and therefore the idea about the role of inductivity was not actively studied. More than half a century later, Fields \citeyearpar{Fields2014a} stated: ``... nerve impulses are not transmitted through neuronal axons the way electrons are conducted through a copper wire, and the myelin sheath is far more than an insulator.'' This statement probably opened more avenues of research.

Contemporary research has revealed many effects accompanying the propagation of an AP that are similar to those caused by inductivity. For example, Koch \citeyearpar{Koch1984} has demonstrated that the quasi-active membrane shows bandpass-like behaviour with a prominent maximum in its membrane impedance. Kumai \citeyearpar{Kumai2017} explains the AP event related to the viewpoint that ``... the Na$^+$ channel is composed of resistance, inductance, and capacitance elements, where the inductance element would function in controlling the Na$^+$ channel activity, and the capacitance element, the K$^+$ channel activity remotely.'' Mosgaard et al. \citeyearpar{Mosgaard2015a} argue that ``the appearance of inductive behaviour in the nerve membrane as well as in pure lipid membranes may arise from the voltage and time dependence of membrane capacitance and membrane conductance.''  Such analysis has lead to the proposal to use the notion of ``phenomenological inductance'' \citep{Mauro1970,Koch1984,Scott2002}. However, as stressed by Scott \citeyearpar{Scott2002}, ``... phenomenological inductance has nothing to do with storage of magnetic field energy, ...''. 

Recently, a paper by Wang et al. \citeyearpar{Wang2021} about inductance as ``a missing piece of neuroscience'' gave a detailed overview of the role of inductivity in myelinated neurons. They attributed the effects of inductivity to the ``equivalent circuit''. It means that there is a kind of biological structure that can store the energy in a non-electrical form and might be related to various structural elements like the membrane and the cytoskeleton \citep{Wang2021} or ion channels \citep{Hodgkin1952}. As a result, ``the neural signal is an energy pulse containing electrical, magnetic, and mechanical components'' \citep{Wang2021}.

Based on these arguments, it might be purposeful to return to basics and not neglect the inductivity when deriving the governing equation for an AP. This is actually done by Lieberstein \citeyearpar{Lieberstein1967}. His model is the following: 
\begin{align}
	&-\frac{\partial i_a}{\partial x}=2\pi a\cdot I+\pi a^2C_a\frac{\partial V}{\partial t},\\
	&-\frac{\partial V}{\partial x}=ri_a+\frac{L}{\pi a^2}\frac{\partial i_a}{\partial t},
\end{align}
where $a$ is the axon radius, $I$ is the membrane current given by Eq.~\eqref{MembCurrent}, $C_a$ axon self-capacitance per unit area per unit length, $r$ is the axon resistance per unit length,$i_a$ is line axon current (along the axon) and $L$ is axon specific self-inductance.

\section{Myelination}

Recently, Tamm et al. \citeyearpar{Tamm2026} have proposed a mathematical model for an AP in myelinated nerve fibres. In this model, the geometry of the myelin sheath is taken into account by two parameters: the g-ratio (the ratio of the inner-to-outer diameter of a myelinated axon) and the $\mu$-ratio ( the ratio of the length of the myelin sheath and the node of Ranvier). Relating to studies of Wang et al. \citeyearpar{Wang2021} about the role of inductance in understanding the process in myelinated nerve fibres, the model of Lieberstein \citeyearpar{Lieberstein1967} is taken as a basic one (see Section~6 ). Leaving aside the details about the structure of myelinated axons \citep{Wang2021,Tamm2026}, the model describing the propagation of the AP is written in the form: 
\begin{align}
	\label{MyelinAPeq1}
	&\frac{\partial V}{\partial t}+\Phi\left[(1+\gamma\mu)\cdot\frac{\partial i_a}{\partial x}+2\pi a\cdot I\right]=0,\\
	&\frac{\partial i_a}{\partial t}+\frac{\pi a^2}{L}\left[\frac{\partial V}{\partial x}+ri_a\right]=0,\\
	&\Phi=\frac{1}{C_a\pi a^2+2C_m\pi a},\\
	&\mu=\frac{l_2}{l_1},\\
	&I=\hat{g}_\mathrm{K}n^4(V-V_\mathrm{K})+\hat{g}_\mathrm{Na}m^3h(V-V_\mathrm{Na})+\hat{g}_\mathrm{L}(V-V_\mathrm{L}),
	\end{align}
where parameter $\mu$ ($\mu$-ratio) describes the average length of the myelinated section $l_2$ divided by the average length of the node of Ranvier $l_1$ and parameter $\gamma$ is a phenomenological coefficient which determines conduction velocity between adjacent nodes of Ranvier.

Like in Section~6, here the original system of two first-order equations is kept, which permits to explain better the role of the phenomenological parameter $\gamma$. This parameter is needed to determine conduction velocity in the axon under the myelin sheath between the adjacent nodes of Ranvier.  The parameter $\gamma$ has values between 0 and 1. The limit cases mean the following: if $\gamma$ is equal to 0, then the current cannot propagate (physically, it means that the nodes of Ranvier are isolated from each other); if $\gamma$ is equal to 1, then the myelinated part works as a perfect conductor (immediate reaction). The combination $\gamma\mu$ in Eq.~\eqref{MyelinAPeq1} can be interpreted like a generalised dimensionless velocity which accounts phenomenologically for physical effects like diffusion, capacitance, inductance or the influence of cytoskeleton. As demonstrated by Tamm et al. \citeyearpar{Tamm2026}, this model captures the increase in the propagation velocity under the myelin sheath.

\section{Conclusions}

Several phenomenological models proposed to describe the propagation of an AP were briefly presented above. The nerve pulse dynamics is an extremely complex process because many physical and chemical effects are coupled and not all these coupling processes are clearly understood physically. However, the experimentalists have demonstrated many details of the formation and propagation of nervous pulses. It means that we have many observations at our disposal, and the question is how these results must be interpreted to build a coherent theory like Andersen et al. \citeyearpar{Andersen2009} suggested (see Section~3). 

In contemporary neuroscience, the HH model is an ingenious example of how phenomenology has helped to combine observations and physical laws, resulting in their celebrated model, which is now the cornerstone of neuroscience. The idea of describing the role of ion currents (of K$^+$ and Na$^+$ ions) in the AP dynamics has been extremely useful. Certainly, the HH model has evoked many further studies to modify the ion currents. Morris and Lecar \citeyearpar{Morris1981} took, for example, the K$^+$ and Ca$^{2+}$ ions into account, resulting in different rate equations. However, two avenues of research have progressed immensely. Although the general ideas of phenomenology started from Husserl and Heisenberg \citep{Bokulich2006}, the iconic HH model has also evoked attention in the philosophy of science. And what is essential for developing neuroscience is the intensive research of ion channels and ion pumps in order to reveal the physical mechanisms that were described phenomenologically in the HH model. 

Based on observations, the phenomenological description answers the question `how?' but the question `why?' is not answered. This is actually a problem of explanation, and many studies are devoted to the analysis of phenomenology and explanation. Craver \citeyearpar{Craver2006}  cites Hodgkin and Huxley \citeyearpar{Hodgkin1952}: ``The agreement [between the model and the voltage clamp data] must not be taken as evidence that our equations are anything more than an empirical description of the time-course of changes in permeability to sodium and potassium. An equally satisfactory description of the voltage clamp data could no doubt have been achieved with equations of very different form, which would probably have been equally successful in predicting the electrical behaviour of the membrane.'' And later Hodgkin \citeyearpar{Hodgkin1992}  added: ``So we settled for the more pedestrian aim of finding a simple set of mathematical equations which might plausibly represent the movement of electrically charged gating particles.'' In continuum mechanics, Maugin \citeyearpar{Maugin1990} has mentioned the similar idea that the choice of internal (hidden)  variables depends on the researcher. Kaplan and Craver \citeyearpar{Kaplan2011} have used the notion `mechanism-we-know-not-what'.

In his analysis of causality, Bogen \citeyearpar{Bogen2005}  stresses that the scientific community should pay attention to looking for the mechanisms responsible for phenomena described faithfully using phenomenological variables by Hodgkin and Huxley. This is the second avenue of intensive research because the mechanisms of ion transport across biomembranes play an important role not only in neuroscience but also in functioning cells with many biological functions. In this field, much is done by Hille and his co-workers starting in the 60-ies of the last century. Their findings explained the molecular mechanisms of ion transport under various conditions. The book entitled ``Ionic Channels of Excitable Membranes'' \citep{Hille1984}  summarised Hille's studies and was later republished in 1992 and 2001. The present understanding on ion channel field is described in a Handbook \citep{Zheng2015}.  

An example of a phenomenological variable influencing the temperature changes, described in Section~5, needs further analysis from the viewpoint of modelling exo- and endothermic reactions in more detail. The idea of using phenomenological inductance  (Section~6) to characterise several processes in axons may open a fascinating new field of studies. Wang et al. \citeyearpar{Wang2021} call inductance ``a missing piece of neuroscience''.   

The hybrid mathematical models, like those described above, combine the physical laws with observables described phenomenologically. In many cases, these models are useful although they do not answer the question `why?', evoking further studies. For example, although Heisenberg introduced himself such variables in the matrices describing initial and final positions of electrons \citep{Heisenberg1925}, later he stressed the need to understand the physical content of the phenomenon (see Bokulich, \citeyearpar{Bokulich2006}). The question is whether the detailed knowledge about the phenomenon helps to construct a better model, or, in other words, is it possible to cast all the information about a complex phenomenon into mathematical language. Indeed, the HH model, for example, is an excellent description and permits the calculation of the AP with sufficient accuracy. The extensive research and detailed description of ion channels \citep{Hille1984,Zheng2015} have not resulted in a better mathematical model describing the AP to replace the HH model. Nevertheless, a phenomenological approach may lead us to further research, i.e., to look for missing pieces. 

\section*{Acknowledgements}
\noindent
This research was supported by the Estonian Research Council (PRG 1227). J\"uri Engelbrecht acknowledges the support from the Estonian Academy of
Sciences.


\end{document}